\documentclass[12pt]{article}
\usepackage{fancyhdr}

\usepackage{mathrsfs}
\usepackage[T1]{fontenc}
\usepackage{mathpazo}
\usepackage{setspace}
\usepackage{amsfonts}
\usepackage{amssymb}
\usepackage{amsmath}
\usepackage{epsfig}
\usepackage{latexsym}
\usepackage{color}
\usepackage{graphicx}
\usepackage{nicefrac}
\usepackage[latin1]{inputenc}
\usepackage{slashed}
\usepackage{multirow}
\usepackage{comment}
\usepackage{soul}
\usepackage{hyperref}
\usepackage[nosort]{cite}
\usepackage{datetime}

\usepackage[titletoc]{appendix}

\def\hybrid{\topmargin -20pt    \oddsidemargin 0pt
        \headheight 0pt \headsep 0pt
        \textwidth 6.25in       
        \textheight 9.25in       
        \marginparwidth .875in
        \parskip 5pt plus 1pt   \jot = 1.5ex}

\hybrid

\def\baselinestretch{1.2}

\catcode`\@=11

\def\marginnote#1{}
\newcount\hour
\newcount\minute
\newtoks\amorpm
\hour=\time\divide\hour by60
\minute=\time{\multiply\hour by60 \global\advance\minute by-\hour}
\edef\standardtime{{\ifnum\hour<12 \global\amorpm={am}%
        \else\global\amorpm={pm}\advance\hour by-12 \fi
        \ifnum\hour=0 \hour=12 \fi
        \number\hour:\ifnum\minute<10 0\fi\number\minute\the\amorpm}}
\edef\militarytime{\number\hour:\ifnum\minute<10 0\fi\number\minute}

\def\draftlabel#1{{\@bsphack\if@filesw {\let\thepage\relax
   \xdef\@gtempa{\write\@auxout{\string
      \newlabel{#1}{{\@currentlabel}{\thepage}}}}}\@gtempa
   \if@nobreak \ifvmode\nobreak\fi\fi\fi\@esphack}
        \gdef\@eqnlabel{#1}}
\def\@eqnlabel{}
\def\@vacuum{}
\def\draftmarginnote#1{\marginpar{\raggedright\scriptsize\tt#1}}

\def\draft{\oddsidemargin -.5truein
        \def\@oddfoot{\sl preliminary draft \hfil
        \rm\thepage\hfil\sl\today\quad\militarytime}
        \let\@evenfoot\@oddfoot \overfullrule 3pt
        \let\label=\draftlabel
        \let\marginnote=\draftmarginnote
   \def\@eqnnum{(\theequation)\rlap{\kern\marginparsep\tt\@eqnlabel}%
\global\let\@eqnlabel\@vacuum}  }

\def\preprint{\twocolumn\sloppy\flushbottom\parindent 2em
        \leftmargini 2em\leftmarginv .5em\leftmarginvi .5em
        \oddsidemargin -.5in    \evensidemargin -.5in
        \columnsep .4in \footheight 0pt
        \textwidth 10.in        \topmargin  -.4in
        \headheight 12pt \topskip .4in
        \textheight 6.9in \footskip 0pt
        \def\@oddhead{\thepage\hfil\addtocounter{page}{1}\thepage}
        \let\@evenhead\@oddhead \def\@oddfoot{} \def\@evenfoot{} }

\def\numberbysection{\@addtoreset{equation}{section}
        \def\theequation{\thesection.\arabic{equation}}}

\def\underline#1{\relax\ifmmode\@@underline#1\else
        $\@@underline{\hbox{#1}}$\relax\fi}

\def\titlepage{\@restonecolfalse\if@twocolumn\@restonecoltrue\onecolumn
     \else \newpage \fi \thispagestyle{empty}\c@page\z@
        \def\thefootnote{\fnsymbol{footnote}} }

\def\endtitlepage{\if@restonecol\twocolumn \else \newpage \fi
        \def\thefootnote{\arabic{footnote}}
        \setcounter{footnote}{0}}  

\catcode`@=12
\relax

\def\figcap{\section*{Figure Captions\markboth
        {FIGURECAPTIONS}{FIGURECAPTIONS}}\list
        {Figure \arabic{enumi}:\hfill}{\settowidth\labelwidth{Figure
999:}
        \leftmargin\labelwidth
        \advance\leftmargin\labelsep\usecounter{enumi}}}
 \relax
\def\tablecap{\section*{Table Captions\markboth
        {TABLECAPTIONS}{TABLECAPTIONS}}\list
        {Table \arabic{enumi}:\hfill}{\settowidth\labelwidth{Table
999:}
        \leftmargin\labelwidth
        \advance\leftmargin\labelsep\usecounter{enumi}}}
 \relax
\def\reflist{\section*{References\markboth
        {REFLIST}{REFLIST}}\list
        {[\arabic{enumi}]\hfill}{\settowidth\labelwidth{[999]}
        \leftmargin\labelwidth
        \advance\leftmargin\labelsep\usecounter{enumi}}}
 \relax
\makeatletter
\newcounter{pubctr}
\def\publist{\@ifnextchar[{\@publist}{\@@publist}}
\def\@publist[#1]{\list
        {[\arabic{pubctr}]\hfill}{\settowidth\labelwidth{[999]}
        \leftmargin\labelwidth
        \advance\leftmargin\labelsep
        \@nmbrlisttrue\def\@listctr{pubctr}
        \setcounter{pubctr}{#1}\addtocounter{pubctr}{-1}}}
\def\@@publist{\list
        {[\arabic{pubctr}]\hfill}{\settowidth\labelwidth{[999]}
        \leftmargin\labelwidth
        \advance\leftmargin\labelsep
        \@nmbrlisttrue\def\@listctr{pubctr}}}
 \relax
\makeatother
\newskip\humongous \humongous=0pt plus 1000pt minus 1000pt

\newif\ifdtup

\relax

\def\be{\begin{equation}}
\def\ee{\end{equation}}
\def\ba{\begin{eqnarray}}
\def\ea{\end{eqnarray}}

\def\del{\partial}

\def\r{\rho}
\def\a{\alpha}

\def\b{\beta}

\def\d{\delta}
\def\D{\Delta}

\def\m{\mu}

\def\l{\lambda}
\def\L{\Lambda}
\def\s{\sigma}

\def\cN{{\cal N}}

\def\cL{{\cal L}}

\def\no{\noindent}

\def\qq{\qquad}

\def\IR{\relax{\rm I\kern-.18em R}}

\def \ov {\over}

\def\IR{\relax{\rm I\kern-.18em R}}
\def\IL{\relax{\rm I\kern-.18em L}}

\def\inv{^{\raise.15ex\hbox{${\scriptscriptstyle -}$}\kern-.05em 1}}

\def\cL{{\cal L}}

\def\Tr{{\rm Tr}}

\begin{document}

\renewcommand{\theequation}{\thesection.\arabic{equation}}
\csname @addtoreset\endcsname{equation}{section}

\newcommand{\beq}{\begin{equation}}
\newcommand{\eeq}[1]{\label{#1}\end{equation}}
\newcommand{\ber}{\begin{equation}}
\newcommand{\eer}[1]{\label{#1}\end{equation}}
\newcommand{\eqn}[1]{(\ref{#1})}
\begin{titlepage}
\begin{center}


${}$
\vskip .2 in

{\large\bf The most general $\l$-deformation of CFTs and integrability 
}

\vskip 0.4in

{\bf George Georgiou}\ \ and\ \ {\bf Konstantinos Sfetsos}
\vskip 0.15in

 {\em
Department of Nuclear and Particle Physics,\\
Faculty of Physics, National and Kapodistrian University of Athens,\\
Athens 15784, Greece\\
}
\vskip 0.12in

{\footnotesize \texttt georgiou@inp.demokritos.gr, ksfetsos@phys.uoa.gr}


\vskip .5in
\end{center}

\centerline{\bf Abstract}

\no
We show that the CFT with symmetry group $G_{k_1}\times G_{k_2}\times \cdots \times G_{k_n}$ consisting of WZW models based on the same group $G$, but at arbitrary integer levels, admits an integrable deformation depending on $2(n-1)$ continuous parameters.
We derive the all-loop effective action of the deformed theory and prove integrability.
We also calculate the exact in the deformation parameters RG flow equations which can be put in a particularly simple compact form.
This allows a full determination and classification of the fixed points of the RG flow, in particular those that
are IR stable.
The models under consideration provide concrete realizations of  integrable flows between CFTs.
We also consider non-Abelian T-duality type limits.

\vskip .4in
\noindent
\end{titlepage}
\vfill
\eject

\newpage

\tableofcontents

\noindent

\def\baselinestretch{1.2}
\baselineskip 20 pt
\noindent


\setcounter{equation}{0}
\section{Introduction }

Integrability is one of the key features for obtaining exact results in quantum field theory (QFT). The most well-known example where integrability
was greatly exploited is that of the maximally supersymmetric gauge theory in four dimensions, that is $\mathcal N=4$ SYM. Thanks to the AdS/CFT correspondence \cite{Maldacena:1997re}
the dynamics of $\mathcal N=4$ SYM can be translated to those of an integrable two-dimensional non-linear $\s$-model evading  thus certain no-go theorems which forbid integrable theories in more than two dimensions.
A variety of integrability-based techniques, from the asymptotic Bethe ansatz \cite{Staudacher:2004tk} to the thermodynamic Bethe ansatz \cite{Ambjorn:2005wa} and the Y-system \cite{Gromov:2009tv}, 
were employed in order to determine the planar anomalous dimensions of gauge invariant operators for all values of the't Hooft coupling $\l=g_{YM}^2 N$.  For further developments on integrability and  the AdS/CFT correspondence see \cite{Beisert:2010jr} and references therein.

It is an intriguing endeavour  to formulate new examples of
gauge/gravity dualities where although supersymmetry may be broken the integrability of the theories involved will be maintained.
The comments above make it clear that  one starting point could be  new integrable two-dimensional field theories which can serve as the seed for such constructions.
One such direction which has attracted attention lately is the $\l$-deformed 
models of \cite{Sfetsos:2013wia},
as well as generalisations in many directions \cite{Georgiou:2017oly,Georgiou:2016urf,Georgiou:2017jfi,Georgiou:2018hpd}.

The aforementioned constructions provide an effective and rather effortless method for obtaining exact results in a general class of two-dimensional QFTs.
The starting point of  the construction is certain conformal field theories (CFTs)
of the WZW type perturbed by current bilinear operators with the currents belonging either to the same and/or different groups. The essence of the method relies on the construction of the corresponding all-loop effective actions
for the deformed theories by a gauging procedure \cite{Sfetsos:2013wia,Georgiou:2017oly,Georgiou:2016urf,Georgiou:2017jfi,Georgiou:2018hpd}.
These effective actions possess non-perturbative symmetries in the space of couplings 
\cite{Itsios:2014lca,Sfetsos:2014jfa,Georgiou:2017jfi,Georgiou:2018hpd}.\footnote{ For isotropic $\l$-deformations the simplest such non-perturbative symmetry was found initially in \cite{Kutasov:1989aw} using path integral arguments.}

Combining low-order perturbation theory with the manifestation of the aforementioned non-perturbative  symmetries at the level of correlators one is able  to derive exact expressions for them.
This program was initiated and implemented in a series of papers in which exact expressions were obtained for many observables. In particular, when gravitational techniques were also implemented the exact in the deformation parameters 
$\b$-functions were found \cite{Itsios:2014lca,Sfetsos:2014jfa,Georgiou:2017jfi,Georgiou:2017aei},
with the most general case of anisotropic $\l$-deformations and different WZW levels found in 
\cite{Sagkrioti:2018rwg}.\footnote{The exact $\b$-functions for CFTs deformed by operators bilinear in currents and for isotropic cases have been obtained in the past either by field theoretical methods (resummation of the perturbation series or  the background field method) \cite{Kutasov:1989dt,Gerganov:2000mt,LeClair:2001yp,Appadu:2015nfa}.
The results are in complete agreement.} In addition, anomalous dimensions of current \cite{Georgiou:2015nka} and primary operators \cite{Georgiou:2016iom,Georgiou:2016zyo}, as well as  three-point correlators involving currents and/or primary operators \cite{Georgiou:2016iom,Georgiou:2016zyo} were calculated for the models of \cite{Sfetsos:2013wia,Georgiou:2017oly,Georgiou:2016urf,Georgiou:2017jfi}. Furthermore, the computation of the $C$-function of Zamolodchikov \cite{Zamolodchikov:1986gt}, exactly in the deformation
parameter for the case of isotropic perturbations but to leading order in $k$, was performed in \cite{c-function:2018}
and further generalized for anisotropic $\l$-deformations in \cite{Sagkrioti:2018abh}. 

\no
In a parallel development
of particular importance  the extension to the case
where the unperturbed CFT of a single WZW model is replaced by a coset CFT was considered in \cite{Sfetsos:2013wia,Hollowood:2014rla,Hollowood:2014qma,Sfetsos:2017sep}. The deformed theory was found to be integrable when the coset is chosen to be a symmetric space \cite{Hollowood:2014rla}.
Subsequently, the corresponding analysis for the case of supergroups was considered in \cite{Hollowood:2014qma}.
Although integrability has not been an essential ingredient  in the computation
of the $\b$-functions
and of the operators anomalous dimensions, in the case of isotropic deformations the above models have been demonstrated to be integrable \cite{Sfetsos:2013wia,Hollowood:2014rla,Hollowood:2014qma,Sfetsos:2017sep,Georgiou:2016urf,Georgiou:2017jfi}. For the special case of the isotropic deformation based on the $SU(2)$ group the model has been initially proven to be integrable in \cite{Balog:1993es}.
Furthermore, integrability
was shown to persist in some other cases with more deformation parameters \cite{Sfetsos:2014lla,Sfetsos:2015nya}.
In addition, certain deformed models of low dimensionality have been embedded to type-IIA or type-IIB supergravity \cite{Sfetsos:2014cea,Demulder:2015lva,Borsato:2016zcf,Chervonyi:2016ajp}.

An interesting relation between $\l$-deformations and $\eta$-deformations for group and coset spaces was discussed in  \cite{Vicedo:2015pna,Hoare:2015gda}, \cite{Sfetsos:2015nya,Klimcik:2015gba,Klimcik:2016rov,Hoare:2018ebg}. In particular, the  $\l$-deformed models are related via Poisson-Lie T-duality,
which has been introduced for group spaces in  \cite{KS95a} and extended for coset spaces in
\cite{Sfetsos:1999zm}, and appropriate analytic continuations to the $\eta$-deformed models. The latter were introduced in \cite{Klimcik:2002zj,Klimcik:2008eq,Klimcik:2014}
and \cite{Delduc:2013fga,Delduc:2013qra,Arutyunov:2013ega} for group and coset spaces, respectively. Moreover, the dynamics of scalar fields in some $\l$-deformed geometries corresponding to
coset CFTs has been discussed in \cite{Lunin:2018vsn} while the relation to Chern-Simons theories was discussed
in \cite{Schmidtt:2017ngw}. Finally,  D-branes regarded as integrable boundary configurations  were introduced in the context of $\lambda$-deformations in
\cite{Driezen:2018glg}.

The models presented in \cite{Georgiou:2017jfi,Georgiou:2018hpd} have several virtues. They provide concrete realizations of  flows between exact CFTs in which the Lagrangian of the theory is known all the way through from the UV to the IR fixed points.
Integrability is preserved in the entire flow and therefore these flows are called integrable. Furthermore, the construction of  \cite{Georgiou:2018hpd} provides the first example in which self- and mutual-interactions between different WZW models are 
present already at leading order in the deformation parameters. In this work
we continue this line of research and construct integrable multi-parameter deformations of CFTs whose Lagrangian and RG flow equations are known to all-orders in perturbation theory. The integrable sector will belong to the most general $\l$-deformed 
model which we construct.  In this all possible self- and mutual-interactions are present.

\no
The plan of the present paper is as follows:
In section 2, we construct the all loop effective action
of a general class of models whose UV Lagrangian is the sum of an arbitrary number $n$ of WZW models based on the same group $G$ but at different levels.
In general our models depend on $n^2$ general coupling matrices.
In section 3, we consider a consistent truncation of our models that depend on $2(n-1)$ couplings,  identify the non-perturbative symmetries in the space of couplings $\hat \l_{i1}$ and $\hat \l{ni}$ and show that  he theory is classically integrable by finding the appropriate Lax connection.
In section 4, we calculate the exact in the deformation parameters $\b$-functions of our models which can be cast in a particularly simple and compact form.
Subsequently, we determine and classify the fixed points of the RG flow in particular those that
are IR stable. Finally, in section 5 we consider non-Abelian T-duality type limits in the case when the theory is integrable. In the last section we present our conclusions.

\section{Generalities:  Lagrangian and equations of motion}

In this section we construct the effective actions of our model and derive the corresponding equations of motion.

Our starting point is to consider group elements $g_i$, $i=1,2,\dots ,n$ in a semi-simple group $G$ and the corresponding WZW model actions $S_{k_i}(g_i)$ at levels $k_1,k_2, \dots , k_n$.
We add to this the action of $n$ PCMs which are both self and
mutually interacting and are constructed using $n$ group elements $\tilde g_i$, $i=1,2,\dots,n$
belonging in the same group $G$. Namely, we have that
\be
\label{oractt}
\begin{split}
& S_{k_i,E_i}(g_i,\tilde g_i)= \sum_{i=1}^nS_{k_i}(g_i)
-{1\ov \pi}\int d^2\s\ \big(\tilde g_i^{-1}\del_+\tilde g_i\big)_a  E_{ij}^{ab} \big(\tilde g_j^{-1} \del_-\tilde g_j\big)_b\ ,
\end{split}
\ee
where the $E_{ij}$, $i,j=1,2,\dots,n$ are generic coupling matrices and the indices
$a,b=1,2,\dots, \dim(G)$.
In the spirit of \cite{Sfetsos:2013wia,Georgiou:2016urf,Georgiou:2018hpd} we gauge the global
symmetry acting on the group elements as follows:
$g_i\to \L_i^{-1} g_i\L_i$ and $\tilde g_i\to \L_i^{-1} \tilde g_i$, $i=1,2,\dots,n$. The resulting
gauge invariant action reads
\ba
\label{gauacc}
S_{k_i,E_i}(g_i,\tilde g_i, A_\pm^{(i)})  = \sum_{i=1}^nS_{k_i}(g_i,A_\pm^{(i)})
-{1\ov \pi}\sum_{i,j=1}^n \int d^2\s\ \Big(\tilde g_i^{-1}\tilde D_+\tilde g_i\Big)_a  E_{ij}^{ab} \Big(\tilde g_j^{-1} \tilde D_-\tilde g_j\Big)_b\,,
\ea
where $S_{k_i}(g_i,A_\pm^{(i)})$ is the standard gauged WZW action
\be
\begin{split}
&  S_{k_i}(g_i,A_\pm^{(i)}) = S_{k_i}(g_i)
+{k_i\ov \pi} \int d^2\s \ \Tr \big(A_-^{(i)} \del_+ g_i g_i^{-1}   - A_+^{(i)} g_i^{-1} \del_- g_i
\\
& \qq\qq\qq + A_-^{(i)} g_i A_+^{(i)} g_i^{-1}-A_-^{(i)}A_+^{(i)}\big) \ ,
\end{split}
\ee
where we have suppressed the group indices.
Furthermore, the covariant derivatives were defined as
$\tilde D_\pm \tilde g_i= (\del_\pm -A_\pm^{(i)}) \tilde g_i$.
One may now fix the gauge in \eqn{gauacc} choosing $\tilde g_i=\mathbb{1}$ to arrive at the
following action
\be
\begin{split}
&  S_{k_i,\l_{ij}}(g_i, A_\pm^{(i)})  = \sum_{i=1}^n S_{k_i}(g_i)
+ {k_i\ov \pi} \int d^2\s \ \Tr \big(A_-^{(i)} \del_+ g_i g_i^{-1}   - A_+^{(i)} g_i^{-1} \del_- g_i
\\
& \qq\qq\qq\ + A_-^{(i)} g_i A_+^{(i)} g_i^{-1}\big)
-{1\ov \pi}\sum_{i,j=1}^n \int d^2\s\ \sqrt{k_i k_j}A_+^{(i)} \l^{-1}_{ij}A_-^{(j)}\,,
\label{gaufix}
 \end{split}
\ee
where we have redefined the coupling matrices appearing in the PCM models
as\footnote{To compare with the corresponding actions in \cite{Georgiou:2018hpd} (see eqs. (2.1) and (2.4)) corresponding to two
WZW models self- and mutually interacting we have
\be
\nonumber
\begin{split}
& E_{11}=E_1\ ,\quad E_{22}=E_4\ ,\quad E_{12}=E_2\ ,\quad E_{21}=E_3\ ,
\\
&
\l_{11} = \sqrt{k_1\ov k_2} \l_4\ ,\quad  \l_{22} = \sqrt{k_2\ov k_1} \l_3\ ,\quad
\l_{12} = \l_2\ ,\quad \l_{21} = \l_1\ .
\end{split}
\ee
}
\be
\begin{split}
& \sqrt{k_ik_j}\ \l_{ij}^{-1}= E_{ij} +k_i \d_{ij}\ .
\end{split}
\ee
In order to obtain the $\s$-model and to show integrability in the next section
we should integrate out the gauge fields. These are not dynamical and appear only quadratically
in the action. We cast the corresponding equations
of motion into the following particularly convenient form
\be
\label{con1}
g_i^{-1} D_- g_i=\Bigg(\d _{ij}-\l _{ij}^{-1}\sqrt{{k_j} \ov {k_i}}\Bigg)A_-^{(j)}\ ,\quad
D_+g_i g_i^{-1}=-\Bigg(\d _{ij}-\l^{-T} _{ji} \sqrt{{k_j} \ov {k_i}}\Bigg)A_+^{(j)}\ ,
\ee
where as usual we have defined  $D_\pm g_i=\partial_\pm g_i-[A_{\pm}^{(i)},g_i]$
and where we note that the transpose, i.e. in $\l^{-T}_{ij}$, as well as the inverses
refer only to the suppressed group indices and not on the space of couplings with indices $i,j$.

\no
Varying with respect to the group elements one obtains
\be
\label{g-eom}
\begin{split}
& D_-(D_+g_i g_i^{-1})=F_{+-}^{(i)}\quad \Longleftrightarrow
\quad D_+(g_i^{-1}D_-g_i)=F_{+-}^{(i)} \ ,
\\
&  F_{+-}^{(i)}=\del_+ A_-^{(i)}-\del_- A_+^{(i)}-[A_+^{(i)},A_-^{(i)}]\ .
\end{split}
\ee
To proceed we find it convenient to rescale the gauge fields and define
\be
\tilde A^{(i)}_\pm = \sqrt{k_i} A^{(i)}_\pm \ .
\ee
Using them and substituting \eqn{con1} in \eqn{g-eom} we obtain after some algebra that
 \be
\label{g-eom1}
\begin{split}
& \l^{-1}_{ij}\del_+ \tilde A_-^{(j)} - \del_- \tilde A_+^{(i)}={1\ov \sqrt{k_i}}
[\tilde A_+^{(i)},  \l^{-1}_{ij} \tilde A_-^{(j)}]\ ,
\\
&\del_+ \tilde A_-^{(i)} - \l^{-T}_{ji} \del_-  \tilde A_+^{(j)}={1\ov \sqrt{k_i}}
[\l^{-T}_{ji} \tilde A_+^{(j)},  \tilde A_-^{(i)}]  \ .
\end{split}
\ee
To present the $\s$-model action we introduce representation Hermitian
matrices $t_a$ and define
\be
J_{+}^a=-i \, \Tr (t^a \del_+ g g^{-1})\ ,\quad
J_{-}^a=-i \, \Tr( t^a g^{-1}\del_- g)\ ,\quad D^{ab}=\Tr (t^a g t^b g^{-1})\ .
\ee
These will be computed for the particular group elements $g_i$ in which case
an extra index $i$ will be inserted in the above quantities. We also define the diagonal in the
coupling space matrix
\be
{\cal D}_{ij}= D_i \d _{ij}\ .
\ee
Then we may solve for the redefined gauge fields using \eqn{con1} to obtain
\be
\tilde A^{(i)}_+=\big(\l^{-T}-{\cal D}\big)^{-1}_{ij}   \sqrt{k_j} J_{j+}\ ,\qq
\tilde A^{(i)}_-=-\big(\l^{-1}-{\cal D}^T\big)^{-1}_{ij} \sqrt{k_j} J_{j-}\ .
\ee
Note that the entries of the matrices $\l^{-T}-{\cal D}$ and $\l^{-1}-{\cal D}^T$
are themselves matrices in the group $G$.
Thus, their inversion is to be understood as an inversion in the space of different models keeping in mind that their entries are non-commutative objects.

\no
Finally, substituting the values for the gauge fields  in the action \eqref{gaufix} we obtain the following $\s$-model
\be
\begin{split}
&  S_{k_i,\l}(g_i)  = \sum_{i=1}^nS_{k_i}(g_i)
+{1\ov \pi}\sum_{i,j=1}^n\int d^2\s\ \sqrt{k_i }J_{i+}\,\Big(\l^{-1}-{\cal D}^T\Big)^{-1}_{ij}\, \sqrt{k_j }
J_{-j}\ .
\label{s-modelgen}
 \end{split}
\ee
This is a particularly compact form resembling the single $\l$-deformed model of \cite{Sfetsos:2013wia}.
The above action encompasses all previous ones for $\l$-deformed models.  
Furthermore it is the most general action that can be construct using the same group $G$ for all couplings. A further generalization in which a different group is associated to each WZW model
 can be constructed straightforwardly.

For small entries in the matrices $\l_{ij}$ the action is
\be
\begin{split}
&  S_{k_i,\l}(g_i)  = \sum_{i=1}^nS_{k_i}(g_i)
+{1\ov \pi}\sum_{i,j=1}^n\int d^2\s\ \sqrt{k_i }J_{i+}\,\big(\l^{-1}\big)^{-1}_{ij}\, \sqrt{k_j }
J_{-j}+ \cdots \ .
\label{s-modelgenlim}
 \end{split}
\ee
Note that $\big(\l^{-1}\big)^{-1}_{ij} \neq \l_{ij}$ since the inverse in $\l^{-1}$
is taken in the group space, and the dots denote subleading terms in the small $\l$-expansion.

\no
In the next section the inversion of the above matrices in the coupling space will be done
explicitly for the case of integrable models.

\section{Integrable deformations}

 In this section we will attempt to answer the following question. For which choices of the matrix $E_{ij}^{ab}$ appearing in \eqref{oractt} is the theory described by \eqn{s-modelgen} integrable?
 To the best of our knowledge we are lacking a general answer to this question even for the case of a single group $G$.
 Nevertheless, it has been proven that there are several cases where these theories are integrable for specific choices of the couplings matrices.
Among the single $\l$-deformations the first one is the case of the isotropic $\l$-deformation, that is when $\l_{ab}=\l \d_{ab}$ \cite{Sfetsos:2013wia}, a second one is the case where the matrix
 $E$ is of the form $E=\frac{1}{t} (1-\eta R)^{-1}$, where the matrix $R$ stisfies the modified Yang-Baxter equation \cite{Klimcik:2008eq,Klimcik:2014}. Another case is that where instead of the group $G$ one has a coset with the coset being a symmetric space
 \cite{Sfetsos:2013wia,Hollowood:2014rla}.\footnote{For the case of the anisotropic ${\l}$-deformed $SU(2)$ model integrability was shown in \cite{Sfetsos:2014lla}.} Furthermore, integrability has been shown for the models of
\cite{Georgiou:2017jfi,Georgiou:2016urf} and \cite{Georgiou:2018hpd}
 representing particular cases of self- and mutual-interactions of current algebras based on WZW models.

\subsection{A truncation of our models}
In what follows, we will show that the theory \eqn{s-modelgen}
is integrable in the case where the matrices $\l_{ij}$ is of the following form
\be
\label{lambda}
\begin{split}
&\l^{-1}_{i1}\neq 0\ , \quad i=1,2,\dots ,n\!-\!1\ ,\qq
\l^{-1}_{nj}\neq 0 \ ,
\quad j=2,3,\dots ,n\ ,
\\
&\l^{-1}_{ij}= 0\,\,\,{\rm for\, all\, other\, entries}\ .
 \end{split}
\ee
Hence, the corresponding coupling matrices reads
\be
\label{lambda22}
\begin{split}
\l^{-1}_{ij} =
\left(       \begin{array}{cccc}
            \l^{-1}_{11} & 0 &  \cdots  &0  \\
             \l^{-1}_{21} & 0 &\cdots  & 0\\
              \vdots  & \vdots  & \ddots & \vdots \\
              \l^{-1}_{(n-1)1}  & 0 & \cdots &  0\\
               0 & \l^{-1}_{n2} & \cdots  & \l^{-1}_{nn}\\
           \end{array}
         \right)\  .
\end{split}
\ee
As a result it will be shown that one obtains an integrable deformation of the $G_{k_1}\times G_{k_2}\times \cdots \times G_{k_n}$ depending on $2(n-1)$ parameters. This provides a partial answer to the question posed in the
beginning of this section. We stress that by turning on just one more coupling, in addition to those appearing in \eqn{lambda22}, it will generate other couplings through quantum corrections.
Then, the theory will  most likely cease to be integrable. However, we have no proof that \eqn{lambda22}
exhausts all integrable cases among our general class of models.

\no
Turning to the general action \eqn{s-modelgen}, this after using \eqn{lambda} and explicitly inverting the relevant matrix
in the coupling space we find that
\be
\begin{split}
&  S_{k_i,\l}(g_i)  = \sum_{i=1}^nS_{k_i}(g_i)
+{k_1\ov \pi}\int d^2\s\ J_{1+}\,(\l^{-1}_{11}-D_1^T)^{-1}\,  J_{1-}
\\
&\qq\quad +{k_n\ov \pi}\int d^2\s\ J_{n+}\, (\l^{-1}_{nn}-D_n^T)^{-1}\,  J_{n-}
-{1\ov \pi}\sum_{i=2}^{n-1} k_i\int d^2\s\ J_{+i}\,D_i\,  J_{i-}
 \\
&\qq\quad + {1\ov \pi} \sum_{i=2}^{n-1} \sqrt{k_1 k_i} \int d^2\s\ J_{i+}\,D_i\ \l^{-1}_{i1}
(\l^{-1}_{11}-D_1^T)^{-1}\,  J_{1-}
\\
&\qq\quad + {1\ov \pi}\sum_{i=2}^{n-1} \sqrt{k_n k_i} \int d^2\s\  J_{n+}
(\l^{-1}_{nn}-D_n^T)^{-1}\,\,\l^{-1}_{ni} D_i\  J_{i-}
\\
&\qq\quad -{1\ov \pi} \sum_{i=2}^{n-1} \sqrt{k_n k_1}\int d^2\s\ J_{n+}
(\l^{-1}_{nn}-D_n^T)^{-1}\,\,\l^{-1}_{ni}\, D_i \,\l^{-1}_{i1}\  (\l^{-1}_{11}-D_1^T)^{-1} J_{1-}\ .
\label{s-modelinter}
 \end{split}
 \ee

\no
From this point on we will focus for simplicity on the case where the couplings are taken isotropic in the group space, that is
\be
\l_{ij}^{ab}=\d^{ab} \l_{ij} \ .
\ee
One now encounters the following problem. By combining $S_{k_i}(g_i),\,i=2,\dots ,n-1$ with the last term in the second line of \eqn{s-modelinter} we obtain
a sum of WZW model actions $S_{-k_i}(g_i^{-1}),\,i=2,\dots ,n-1$ with negative signature 
(this is also true for the PCM part of the action \eqn{oractt} before the gauging procedure is performed).
As was pointed out in the case of two interacting WZW models in \cite{Georgiou:2018hpd} to remedy this situation one can perform the following redefinitions of the couplings and analytic continuation in the following specified order: First define the hatted couplings
\be
\label{redeff1}
 \hat \l_{i1}=  \l_{11}\l^{-1}_{i1} \sqrt{{k_1 \ov k_i}}\ ,\qquad  \hat \l_{ni}=  \l_{nn}\l^{-1}_{ni} \sqrt{{k_n\ov k_i}}\  ,\qq
\ i=2,3,\dots ,n-1\
\ee
and then flip the signs of levels and invert the corresponding group elements as
\be
\label{redeff2}
 k_i\to -k_i\ ,\qquad   g_i\to g_i^{-1} \ , \qq  i=2,3,\dots ,n-1 \ .
\ee
Then the action \eqn{s-modelinter} becomes
\be
\begin{split}
&  S_{k_i,\l}(g_i)  = \sum_{i=1}^nS_{k_i}(g_i)
\\
&
\quad\ +{k_1\ov \pi}\int d^2\s\ J_{1+}\,(\l^{-1}_{11}\mathbb{1} -D_1^T)^{-1}\,  J_{1-}
+{k_n\ov \pi}\int d^2\s\ \ J_{n+}\,(\l^{-1}_{nn}\mathbb{1}-D_n^T)^{-1}\, J_{n-}
\\
&
\quad\ +{1\ov\pi} \sum_{i=2}^{n-1}  k_i \int d^2\s\ \l_{11}^{-1}\ \hat \l_{i1}\ J_{i+}
(\l^{-1}_{11}\mathbb{1}-D_1^T)^{-1}\,  J_{1-}
\\
&
\quad\ +{1\ov \pi} \sum_{i=2}^{n-1}   k_i  \int d^2\s\ \l_{nn}^{-1}\ \hat \l_{ni}\ J_{n+}  (\l^{-1}_{nn}\mathbb{1} -D_n^T)^{-1}\,  J_{i-}
\\
&\quad\
+{1\ov\pi} \sum_{i=2}^{n-1}  k_i  \int d^2\s\ \l_{11}^{-1} \l_{nn}^{-1}\ \hat \l_{ni} \ \hat \l_{i1}\ J_{n+}
(\l^{-1}_{nn}\mathbb{1}-D_n^T)^{-1}\, D_i^T \,  (\l^{-1}_{11}\mathbb{1}-D_1^T)^{-1}J_{1-}\ .
\label{s-modelfin}
 \end{split}
 \ee
This is the final expression for the all-loop effective action of our $\s$-model. In what follows we will prove that it is indeed classically integrable.

\no
Before that, we comment on the constraints imposed by demanding a non-singular $\s$-model of Euclidean signature.
In order to avoid singularities the couplings $\l_{11}$ and $\l_{nn}$ should be such that  $| \l_{11}|,|\l_{nn}|<1$,
the reason being that the matrix $D$ is orthogonal and therefore its eigenvalues lie in the
unit circle. In addition,
similarly to \cite{Georgiou:2018hpd}, it can be shown that the signature of \eqn{s-modelfin} is Euclidean provided that the couplings lie within the ellipsoids defined by
\be\label{Deltas}
\D_1= 1-\l_{11}^2- \sum_{i=2}^{n-1} {k_i\ov k_1} \hat \l_{i1}^2>0
\quad {\rm and} \quad  \D_2= 1-\l_{nn}^2- \sum_{i=2}^{n-1} {k_i\ov k_n} \hat \l_{ni}^2>0\ .
\ee

\no
It is straightforward to show that the action \eqn{s-modelfin} is independently invariant under the following two non-perturbative symmetry operations, as well as under their combination
\be
\begin{split}
&({\rm A}): \quad\l_{11}\to {1\ov \l_{11}}\ ,\quad \hat \l_{i1}\to {\hat \l_{i1}\ov \l_{11}},\quad k_1\to -k_1,\ \quad g_1\to g_1^{-1},\ i=2,\dots,n-1\\
&({\rm B}): \quad\l_{nn}\to {1\ov \l_{nn}}\ ,\quad \hat \l_{ni}\to {\hat \l_{ni}\ov \l_{nn}},\quad k_n\to -k_n,\ \quad g_n\to g_n^{-1},\ i=2,\dots ,n-1\ .
\label{symm}
\end{split}
\ee
This is a generalization of the corresponding symmetry found in \cite{Georgiou:2018hpd}.

\no
Expanding \eqn{s-modelfin} and keeping the linear terms in the couplings one obtains
 \be
\begin{split}
&  S_{k_i,\l}(g_i)  = \sum_{i=1}^nS_{k_i}(g_i)
+{1\ov \pi}\int d^2\s\big( k_1\l_{11}\ J_{1+}^a\, J_{1-}^a +k_n  \l_{nn}\ J_{n+}^a\,
J_{n-}^a\big)
\\
&\qq\qq + { 1\ov \pi}\sum_{i=2}^{n-1} k_i \int d^2\s\ \big( \hat \l_{i1}\ J_{i+}^a  J_{1-}^a+  \hat \l_{ni}\ J_{n+}^a \,  J_{i-} ^a\big)+{\cal O}(\l^2)\ .
\label{s-modellin}
 \end{split}
 \ee
Notice the last line of \eqn{s-modelfin} has disappeared from the small coupling expansion since it is quadratic in the $\l$'s. This term, as well as the full action of
\eqn{s-modelfin} is generated when quantum corrections at arbitrary order in perturbation theory are taken into account.

\no
Note that there is the following interesting truncation of \eqn{s-modelfin} which dramatically further simplifies it.
If we let $\l_{11}=\l_{nn}=0$ then we obtain
\be
\begin{split}
&  S_{k_i,\l}(g_i)  = S_{k_1}(g_1) + S_{k_n}(g_n) +  \sum_{i=2}^{n-1}S_{k_i}(g_i)
\\
&
\qquad\ +{1\ov\pi} \sum_{i=2}^{n-1}  k_i \int d^2\s\ \big(\hat \l_{i1}\ J_{i+}\,  J_{1-}
 +  \hat \l_{ni}\ J_{n+} \,  J_{i-} + \hat \l_{ni} \ \hat \l_{i1}\ J_{n+}
\, D_i^T \, J_{1-}\big)\ .
\label{s-modelfin9}
 \end{split}
 \ee
This action is at most quadratic in the couplings and represents only mutual interactions between the various WZW models 
as depicted. It becomes linear if we make the further truncation $\hat\l_{n1}=0$ or $\hat \l_{i1}=0$. 

\subsection{Proof of integrability}

To prove that \eqn{s-modelfin} is integrable we write the
equations of motion \eqn{g-eom1} for the choice of couplings appearing in \eqn{lambda}. The first equation of \eqn{g-eom1} can be decomposed to
 \be
\label{eom1}
\begin{split}
& \l^{-1}_{i1} \del_+ \tilde A_-^{(1)}-\del_- \tilde A_+^{(i)}={1\ov \sqrt{k_i}}[\tilde A_+^{(i)},  \l^{-1}_{i1} \tilde A_-^{(1)}]\ ,\,\,\, \qquad i=1,2,\dots,n-1\ ,
\\
&\l^{-1}_{ni} \del_+ \tilde A_-^{(i)}-\del_- \tilde A_+^{(n)}={1\ov \sqrt{k_n}}[\tilde A_+^{(n)},  \l^{-1}_{ni} \tilde A_-^{(i)}]\ .
\end{split}
\ee
while the second one to
\be
\label{eom2}
\begin{split}
&\del_+ \tilde A_-^{(1)}- \l^{-1}_{i1} \del_-  \tilde A_+^{(i)}={1\ov \sqrt{k_1}}[ \l^{-1}_{i1} \tilde A_+^{(i)},   \tilde A_-^{(1)}]\ ,
\\
&\del_+ \tilde A_-^{(i)}-\l^{-1}_{ni}  \del_- \tilde A_+^{(n)}={1\ov \sqrt{k_i}}[ \l^{-1}_{ni} \tilde A_+^{(n)},  \tilde A_-^{(i)}]\ ,\,\,\, \qquad i=2,3,\dots ,n\ .
\end{split}
\ee
The first line equations in \eqn{eom1} can be combined with the equation in the first line in \eqn{eom2}  to form a system
depending only on the $ \l^{-1}_{i1}$'s but not on $ \l^{-1}_{ni}$ which furthermore can be used to
solve for the $n$ derivatives of the gauge fields,  $\del_+ \tilde A_-^{(1)}$ and $\del_- \tilde A_+^{(i)},\ i=1,\dots,n-1$.
Similarly, we may use the  equation  in second line of \eqn{eom1} and 
 the equations in the second line in \eqn{eom2} to form a system of equations that depend only on the $ \l^{-1}_{ni}$'s,  but not on the $ \l^{-1}_{i1}$'s.
This allows to solve for the $n$ derivatives of the gauge fields $\del_- \tilde A_+^{(n)}$ and $\del_+ \tilde A_-^{(i)}$, $i=2,\dots ,n$.
The two aforementioned sets of equations are decoupled. We stress that even though in these systems the couplings
$ \l^{-1}_{i1}$  completely disentangled from the $ \l^{-1}_{ni}$'s  this is not the case in the effective action \eqref{s-modelfin} where the last term involves both sets of couplings. 
The reason is that the various gauge fields depend on all group elements $g_i$, $i=1,2,\dots, n$.

\no
Our strategy will be to determine a Lax pair for each set of equations and then show that the charges obtained from the first Lax pair are in involution with those obtained from the second one, thus proving that the theory is integrable.
To this end we define  the rescaled anticommuttators
\be
[\tilde A^{(i)}_+,\tilde A^{(j)}_-]_{*L}={1 \ov \sqrt{k_i}}[\tilde A^{(i)}_+,\tilde A^{(j)}_-]\ ,
\qq
[A^{(i)}_+,A^{(j)}_-]_{*R}={1 \ov \sqrt{k_j}}[A^{(i)}_+,A^{(j)}_-]\ .
\ee
In terms of these redefined
quantities the solution to the first set of equations reads
\be 
\label{sol1}
\begin{split}
&\del_+ \tilde A_-^{(1)}={1\ov d}\sum_{i=1}^{n-1}  (\mu_{i1}-\l_{i1}^{-1})\l_{i1}^{-1}[\tilde A^{(i)}_+,\tilde A^{(1)}_-]_{*L}\ , \ \ \mu_{i1}=\sqrt{{k_i \ov k_1}}\ ,\ \ d=1-\sum_{i=1}^{n-1} \l_{i1}^{-2}\ ,
\\
&\del_- \tilde A_+^{(i)}={\l_{i1}^{-1}\ov d}\Big(-d [\tilde A^{(i)}_+,\tilde A^{(1)}_-]_{*L}  +\sum_{j=1}^{n-1}(\mu_{j1}-\l_{j1}^{-1})\l_{j1}^{-1}[\tilde A^{(j)}_+,\tilde A^{(1)}_-]_{*L}\Big)\ .
\end{split}
\ee
These imply the existence of a Lax pair which is constructed by first assuming that 
this is of the form
\be
\begin{split}
\label{Lax}
\cL^{(1)}_+=\sum_{i=1}^{n-1}c_+^{(i)}(z) \tilde  A_+^{(i)} ,\qquad
\cL^{(1)}_-=z \tilde A_-^{(1)} ,
\end{split}
\ee
where $c_+^{(i)}$  are constants depending
on the matrix of the couplings, on the WZW levels as well as on the spectral parameter.
After substituting into the Lax equation
\be\label{Lax-gen}
\partial _+\cL^{(1)}_--\partial _-\cL^{(1)}_+ -[\cL^{(1)}_+,\cL^{(1)}_-]=0\ ,
\ee
and using \eqn{sol1} one obtains a system $n$ equations which can be solved for unknowns $c_+^{(i)}, \ i=1,\dots ,n-1$
arising by equating to zero the coefficients of  $[\tilde A^{(i)}_+,\tilde A^{(1)}_-]_{*L}$. This reads
\be
\begin{split}\label{coeff}
{z \ov d}-{1 \ov d} \sum_{j=1}^{n-1}(c_+^{(j)}\l^{-1}_{j1})+{(\l^{-1}_{i1} -z\sqrt{k_i})\ov \l^{-1}_{i1}(\mu_{i1}-\l^{-1}_{i1}) }c_+^{(i)}=0\ .
\end{split}
\ee
The solution is
\be
\begin{split}\label{c-sol}
&c_+^{(i)}={\l^{-1}_{i1}(\l^{-1}_{i1}-\mu_{i1}) \ov(\l^{-1}_{i1} -z\sqrt{k_i})}\ {z \ov d+d_1}\ ,\ \
i=1,2,\dots ,n-1\ ,\ \  d_1=\sum_{j=1}^{n-1} {\l^{-2}_{j1}(\l^{-1}_{j1}-\mu_{j1} )\ov \l^{-1}_{j1} -z\sqrt{k_j} }\ .
\end{split}
\ee
In conclusion we have shown that the equations of motion \eqn{sol1} imply the existence of a Lax pair from which an infinite tower of conserved charges can be calculated.

In a similar manner, the second system of equations can be solved for the derivatives of the gauge fields as 
\be
\label{sol2}
\begin{split}
&\del_- \tilde A_+^{(n)}={1\ov \hat d}\sum_{i=2}^{n}(\mu_{ni}-\l_{ni}^{-1})\l_{ni}^{-1}[\tilde A^{(n)}_+,\tilde A^{(i)}_-]_{*R}\ , 
\ \  \mu_{ni}=\sqrt{{k_n \ov k_i}}\ ,\ \ \hat d=1-\sum_{i=2}^{n} \l_{ni}^{-2}\ ,
\\
&\del_+ \tilde A_-^{(i)}={\l_{ni}^{-1}\ov \hat d}\Big(-\hat d [\tilde A^{(n)}_+,\tilde A^{(i)}_-]_{*R}  +\sum_{j=2}^{n}(\mu_{nj}-\l_{nj}^{-1})\l_{nj}^{-1}[\tilde A^{(n)}_+,\tilde A^{(j)}_-]_{*R}\Big)\ .
\end{split}
\ee
The corresponding Lax pair will be of the form
\be
\begin{split}
\label{Lax2}
\cL^{(2)}_-=\sum_{i=2}^{n}c_-^{(i)} \tilde  A_-^{(i)} ,\qquad
\cL^{(2)}_+=z \tilde A_+^{(n)} .
\end{split}
\ee
The flatness of this Lax pair is guaranteed when
\be
\begin{split}\label{c-sol2}
&c_-^{(i)}={\l^{-1}_{ni}(\l^{-1}_{ni}-\mu_{ni}) \ov(\l^{-1}_{ni} -z\sqrt{k_i})}\ {z \ov \hat d + \hat d_1}\ ,\quad
i=2,\dots,n\ ,\quad  \hat d_1=\sum_{j=2}^n {\l^{-2}_{nj}(\l^{-1}_{nj}-\mu_{nj} )\ov \l^{-1}_{nj} -z\sqrt{k_j} }\ .
\end{split}
\ee

\no 
The conserved charges obtained from \eqn{Lax} and \eqn{Lax2} are in involution. This is so because any the gauge fields appearing in \eqn{Lax}
have zero Poisson bracket with any of the the gauge fields appearing in \eqn{Lax2}. To see this one may define,
similarly to what was first done in the Hamiltonian treatment of gauged WZW models in \cite{Bowcock}, 
the following dressed currents $\jmath_\pm^{(i)}$ which satisfy the usual
Kac-Moody algebra at level $k_i$
\be
\begin{split}\label{dress-curr}
\jmath_+^{(i)}=D_+g_i g_i^{-1}+A_+^{(i)}-A_-^{(i)}, \qquad \jmath_-^{(i)}=- g_i^{-1}D_-g_i-A_+^{(i)}+A_-^{(i)}.
\end{split}
\ee
Using these definitions and the constraints \eqn{con1}  one obtains
\be
\begin{split}\label{dress-curr}
&\jmath_+^{(1)}=\sum_{i=1}^{n-1} \sqrt{{k_i \ov k_1}}\l^{-1}_{i1} A_+^{(i)}-A_-^{(1)}\ ,
 \\
&\jmath_-^{(i)}=\sqrt{{k_1 \ov k_i}}\l^{-1}_{i1} A_-^{(1)}-A_+^{(i)},\qquad i=1,2, \ldots, n-1\ .
\end{split}
\ee
These can be inverted to express $A_-^{(1)} $ and the $A_+^{(i)}$'s in terms of $\jmath_+^{(1)}$ and 
the $\jmath_-^{(i)}$'s.
In precisely the same way the gauge fields  $A_+^{(n)} $ and $A_-^{(i)}$ can be expressed  in terms of the dressed currents
$\jmath_-^{(1)}$ and $\jmath_+^{(i)}$. Due to the fact that the first set of currents has zero Poisson brackets with those of the second set we conclude that
$\{\cL^{(1)}_\pm,\cL^{(2)}_+\}_{PB}=0$ and as a result the conserved charges obtained from the monodromy matrix involving the first Lax pair are in involution with those obtained from the monodromy matrix involving  the second. This concludes the proof
that the CFT deformed by the $2(n-1)$ deformation parameters $\l^{-1}_{i1}, \ i=1,2, \ldots, n-1$ and $\l^{-1}_{ni}, \ i=2,3, \ldots, n$ is integrable.

A final comment is in order. The careful reader may have noticed that the analysis for integrability has been done using the equations of motion \eqn{eom1} and  \eqn{eom2} in which the analytic continuation discussed in \eqn{redeff2}
has not been applied yet. However, one can easily see that the analytic continuation of \eqn{redeff2} can be straightforwardly be done in the final expressions for the Lax pairs \eqn{Lax} and \eqn{Lax2} once the latter are expressed in terms of the usual gauge fields $A_\pm^{(i)}$ and the couplings $\hat \l_{i1}$ and $\hat \l_{ni}$ appearing in the $\s$-model \eqn{s-modelfin}.
\section{RG flow equations and fixed points}

\subsection{The RG flow equations}
In this section we will first calculate the running of the couplings in the case of the integrable deformations presented
in the previous section. We will need the system of RG flow equations  for the running of the couplings of the anisotropic $\l$-model \cite{Sfetsos:2014jfa}
\be\
\b_{AB}={\mathrm{d}\L_{AB} \ov \mathrm{d}t} = {1\ov 2k} \cN(\L)_{AC}{}^D \cN(\L^T)_{BD}{}^C
\label{dgdgg3} \  ,
\ee
where $t=\ln \m^2$, $\m$ being the energy scale and where
\be
\begin{split}
&
{\cN}(\L)_{AB}{}^c = \left(\L_{AE}\l_{BD}f_{EDF} - f_{ABE} \L_{EF}\right) g^{FC} \ ,
\\
& \tilde g_{AB}=(\mathbb{I}-\L\L^T)_{AB}\ , \qq g_{AB}=(\mathbb{I}-\L^T\L)_{AB}\ .
\end{split}
\ee
In what follows each of the capital indices of \eqref{dgdgg3} will be split in two, that is
 $A=(ia)$, where $i,j,k=1,2,\dots,n-1$ enumerate the different groups while $a,b,c=1,\dots,\dim(G)$ denote as usual
 group indices.
Furthermore, in order to take into account for the different levels $k_i$ of the WZW models, one should set $k=1$ in \eqref{dgdgg3} and rescale the structure constants of each group as $f_{ABC}=f_{(ia)(ib)(ic)}= f_{abc}/\sqrt{k_i}$.
All structure constants that do not have the indices enumerating the different groups equal are set, of course to zero.
Finally, the coupling constants matrix $\l_{AB}$ can be read from \eqn{s-modellin}. It reads
\be\label{lambda2}
\begin{split}
\L_{ij}=
\left(       \begin{array}{cccc}
            \l_{11}& 0 &  \cdots  &0  \\
             \hat \l_{21} \sqrt{{k_2 \ov k_1}} & 0 &\cdots  & 0\\
              \vdots  & \vdots  & \ddots & \vdots \\
              \hat \l_{(n-1)1} \sqrt{{k_{n-1} \ov k_1}} & 0 & \cdots &  0\\
               0 & \hat \l_{n2} \sqrt{{k_2 \ov k_n}} & \cdots  & \l_{nn}\\
           \end{array}
         \right)\  .
\end{split}
\ee
Keeping in mind that the couplings of the first column in \eqn{lambda2} decouple from the ones in the last row of the same equation we focus on the first column and  rename 
$\l_{11}$ by $\l_1$ and similarly $\hat \l_{i1}$ by $\l_{i}$, where $i=2,3,\dots, n-1$. Then the non-zero components of ${\cal N}_{ia, ib}{}^{jc}(\L)$ and  ${\cal N}_{1a, 1b}{}^{ic}(\L^T)$ are
\be
\begin{split}\label{Ns}
&{\cal N}_{ia, ib}{}^{1c}(\L)={ \l_{i} \ov \sqrt{k_1}\D}({k_i\ov k_1} \l_{i}-1) f_{abc}\ , \qquad  i=1,1,\dots, n-1,
\\
&{\cal N}_{1a, ib}{}^{1c}(\L)={ \l_{1} \l_{i} \sqrt{k_i}\ov k_1\D} f_{abc}\ , \qquad  i=2,3,\dots, n-1
\\
&{\cal N}_{ia, jb}{}^{1c}(\L)={  \l_{i} \l_{j}  \sqrt{k_i k_j}\ov k_1^{3/2} \D} f_{abc}\ , \qquad  i\neq j,\quad  i,j=2,3,\ddots, n-1\ ,
\end{split}
\ee
where
\be
\D=1-\sum_{i=1}^{n-1}{k_i\ov k_1}  \l_{i}^2\ ,
\ee
is the same quantity, called $\D_1$ in \eqn{Deltas}, after the renaming of the couplings we mentioned.
Furthermore,
\be
\begin{split}\label{Nts}
&{\cal N}_{1a, 1b}{}^{ic}(\L^T)={\l_{i} \sqrt{k_i}\ov k_1 \D}\big( \l_{i} \D-1+\sum_{j=1}^{n-1}{k_j\ov k_1} \l_{j}^3\big) f_{abc}\ ,\qq i=1,2,\dots, n-1.
\end{split}
\ee
 Notice that 
 ${\cal N}_{ia, jb}{}^{kc}(\L)=-{\cal N}_{jb, ia}{}^{kc}(\L)$ 
 and ${\cal N}_{ia, jb}{}^{kc}(\L^T)=-{\cal N}_{jb, ia}{}^{kc}(\L^T)$.
Using the expressions above we finally obtain for the running of the couplings the following formula
\be
\label{beta-functions}
\b_i = -{c_G\ov 2 k_1}{\l_i(1-\l_i) (\l_i\D - Z)\ov \D^2}\ , \qq i=1,2,\dots , n-1 \ ,
\ee
where 
\be
Z=\sum_{i=1}^{n-1}{k_i\ov k_1}\l_i^2(1-\l_i)\ .
\ee
We note that despite of the privileged r\^ole of $\l_{1}$ in the action \eqn{s-modelfin} (see also \eqn{s-modellin}) the $\b$-functions for all couplings are on equal footing and can be obtained from a single expression \eqn{beta-functions}.
This fact will allow us to fully determine the fixed points of the RG-flow in the next sections.

\no
Finally, similar expressions hold for the running of the couplings $\hat \l_{ni}$. The same is true for the analysis 
following in this section.

\subsection{Fixed points}

We now determine the fixed points of the RG flow equations \eqn{beta-functions}. Each fixed point
belongs to one of the following classes. Each class is characterised by three integers $(n_1,n_2,n_3)$,
obeying the condition $n_1+n_2+n_3=n-1$. By $n_1$ we denote the number of the couplings that are set to zero, that is
$\l_{m_i}=0,\ i=1,2,\dots , n_1$. These couplings can be distributed randomly among the complete set of the $n-1$ couplings.
By $n_2$ we denote the number of the couplings that are set to one, that is
$\l_{p_i}=1,\ i=1,2,\ldots , n_2$. Finally, by $n_3$ we denote the number of couplings that are neither zero nor one, 
i.e. $\l_{q_i}\neq 0,1$, with $i=1,2,\dots, n_3$. By subtracting pairwise the equations in \eqn{beta-functions} it is straightforward to see that all the $\l_{q_i}$
which are neither zero nor one should be equal to each other, that is $\l_{q_i}=\l_*,\ i=1,2,\ldots, n_3$. 
Then the quantity $\D$ becomes
\be
\D=1-{\rho \ov k_1} - {\zeta\ov k_1}  \l_*^2\ ,\qquad \rho=\!
\sum_{i=1}^{n_2}k_{p_j}\ ,\qquad
\zeta=\! \sum_{i=1}^{n_3} k_{q_j}\ .
\ee
As a result the vanishing of the bracket in  \eqn{beta-functions} implies
\be\label{fixed-gen}
\l_*={k_1-\rho \ov \zeta}\ .
\ee
We will assume in the rest of our analysis that $n_3\geqslant 1$ since it turns out that this should be the case 
for the existence of physical IR stable fixed points. 

\no
Let us note that each of the distinct classes of fixed points characterised by $(n_1,n_2,n_3)$,
that obey the condition $n_1+n_2+n_3=n-1$ 
 has $\displaystyle {(n-1)! \ov n_1! n_2! n_3!}$ different members.

\subsection{The stability matrix}

We evaluate the stability matrix for each of the fixed points given by \eqn{fixed-gen}.
This will allow us to determine the relevant and irrelevant directions of each of the fixed points and identify
the fixed points which are IR stable.

Let's split the index $i=(m, a,\a)$ in such a way that
\be
\begin{split}
& \l_m=0\ ,\qquad m=1,2,\dots , n_1\ ,
\\
&  \l_a=1\ ,\qquad a=1,2,\dots,n_2\ ,
\\
& \l_\a=\l_* = {k_1-\r\ov \zeta}\ ,\qquad \a=1,2,\dots,n_3\ ,
\end{split}
\ee
Then at the FPs we have the following relations
\be
\label{ggfDD}
\D ={\zeta\ov k_1} \l_* (1-\l_*)\ ,\qq Z=\l_* \D\ .
\ee
Furthermore, the derivatives of $\D$ and $Z$ evaluated at the FPs  read
\be
\begin{split}
&
\del_m \D = \del_m Z=0
\\
& \del_a \D =-2{k_a\ov k_1}\ ,\qq  \del_a Z=-{k_a\ov k_1}\ ,
\\
& \del_\a \D =-2{k_\a\ov k_1}\l_*\ ,\qq  \del_\a Z={k_\a\ov k_1}\l_*(2-3 \l_*)\ ,
\end{split}
\ee
The non-zero entries of the stability matrix defined as $H_{ij}=\del_j \b_i|_{\rm FP}$ are given by
\be
\label{stabb}
\begin{split}
& H_{mn}= {c_G\ov 2k_1} {\l_*\ov \D}\d_{mn}\ ,\qq H_{ab} = {c_G\ov 2k_1} {1-\l_*\ov \D}\d_{ab}\ ,
\\
& H_{\a\b}= -{c_G\ov 2k_1} {\l_*(1-\l_*)\ov \D^2}\Big(\D \d_{\a\b} - { k_\b\ov k_1}\l_*(2-\l_*)\Big)\ ,
\\
&
 H_{\a a}= -{c_G\ov 2k_1} {\l_*(1-\l_*)\ov \D^2}{k_a\ov k_1} (1-2 \l_*)\ .
 \end{split}
\ee
Note that $H_{\a a}=0$, so that the stability matrix is not a symmetric one.
Clearly the part of this matrix corresponding to the $n_1$ values $\l_m=0$ decouples.
For the eigenvalue problem for the rest of the stability matrix we clearly obtain a matrix of the form
\be
\label{abc0}
\left(
  \begin{array}{cc}
   A &  {\bf 0}  \\
    B & C \\
  \end{array}
\right)= \left(
  \begin{array}{cc}
    \mathbb{1} & {\bf 0}\\
    {\bf 0} & C \\
  \end{array}
\right)  \left(
  \begin{array}{cc}
    A &   {\bf 0}\\
   C^{-1} B& \mathbb{1} \\
  \end{array}
\right)  \ .
 \ee
Specifically, we have the matrix elements
$A_{ab} = H_{ab} - E\d_{ab}$ already in diagonal form with $E$ appearing in the left hand side of
\eqn{abc0} being an eigenvalue of the stability matrix.
The matrix $C_{\a\b}= H_{\a\b}-E \d_{\a\b}$ is of the form $c_1 \d_{\a\b} -c_2 k_\b$ and
as a result its determinant is equal to $c_1^{n_3-1}(c_1-\zeta c_2)$. Furthermore, the second matrix in the 
right hand side of \eqn{abc0} is triangular.
The vanishing of the determinant of the matrix in \eqn{abc0} gives the eigenvalues of the stability matrix, as well as the corresponding degeneracies
\be\label{degen}
\begin{split}
& {\rm Deg}=n_1 :\qq\quad\ \  H_0= {c_G\ov 2k_1} {\l_*\ov \D}\ ,
\\
& {\rm Deg}=n_2 :\qq\quad\ \ H_1= {c_G\ov 2k_1} {1-\l_*\ov \D}\ ,
\\
& {\rm Deg}=n_3-1 :\qq\! H_*=
-{c_G\ov 2k_1} {\l_*(1-\l_*)\ov \D}\ ,
\\
&  {\rm Deg}=1: \qq\qq
H_{1*} = {c_G\ov 2k_1} {\zeta\ov k_1} {\l_*^2(1-\l_*)\ov \D^2}\ .
\end{split}
\ee
These eigenvalues and their degeneracies will be instrumental below in determining the physical IR stable fixed points of
the RG flow. 

\subsubsection{IR stable fixed points}

It turns out that in order to have a Euclidean signature for the metric, the constant $\D$ should be positive leading  
to the following condition
\be
0<\l_*<1 \quad  \Longleftrightarrow\quad \r< k_1< \r +\zeta \ .
\label{hfl01}
\ee
As mentioned before we 
will only consider cases with $n_3\geqslant 1$.
In addition, physical fixed points cannot have $\l_1=1$ since the  $\s$-model action \eqn{s-modelfin} is 
in that case singular.

\no
We are primarily interested in identifying the physical  IR stable points, that is the ones which have all eigenvalues of the stability matrix positive.
The condition \eqn{hfl01} guaranties that $H_0>0$ and $H_1>0$. However, the third eigenvalue in \eqn{degen} is negative, i.e. $H_*<0$ 
and therefore we should necessarily have $n_3=1$ so that this eigenvalue is non-existing. 
Finally, the positivity of the last eigenvalue $H_{1*}$  is also guaranteed  by \eqn{hfl01}.

The condition $n_3=1$ means that there is a single coupling that is not zero or one. 
Keeping in mind that  $\l_1\neq 1$, there are two cases. In the first one let $\l_1=0$ and a single $\l_a=\l_*$. Then also 
$\zeta=k_a$. The condition  \eqn{hfl01} implies that $\r<k_1<\r+ k_a$. 
The second case is when $\l_1=\l_*$. In this case $\l_1=1-{\r/ k_1}$ which holds automatically since
according to \eqn{hfl01}. Also, in this case $\zeta=k_1$. We conclude that there is a multitude of fixed points the number of which depends on the relative ordering of the levels $k_i$. It would be certainly interesting to study the structure and properties of the corresponding
CFTs.

\section{The non-Abelian T-duality limit}

When some of the $\l$'s approach unity then we get a singularity in the manifold.
Then a zoom in procedure maybe applied as in \cite{Sfetsos:2013wia}.
We will not discuss this for the most general action \eqn{s-modelgen} but for
\eqn{s-modelfin} corresponding to integrable models.

\no
In that case case near $\l_{11}=1$ or near $\l_{nn}=1$ we get a singularity in the manifold. However, one may zoom in by taking simultaneously the large $k_1$ and $k_n$ limits
Then we expand for $k_1,k_n\gg 1$ as
\begin{equation}
\begin{split}
& \lambda_{11} =1 - {1\ov 2\zeta_1 k_1}  +\cdots  \ ,
\quad g_1 = \mathbb{I} + {i\ov 2\zeta_1}
 {(v_1)_a t^a \ov k_1} + \cdots  \ ,
\\
&
\lambda_{nn} =1 - {1\ov 2\zeta_n k_n}  +\cdots\ ,
\quad g_n = \mathbb{I} + {i\ov 2\zeta_n}
 {(v_n)_a t^a \ov k_n} + \cdots\ ,
\end{split}
\label{laborio}
\end{equation}
where $\zeta_1 $ and $\zeta_n$  are new coupling parameters.
This leads to
\begin{equation}
\begin{split}
& J_{1\pm}^a ={1\ov 2\zeta_1} {\del_\pm v_1^{a}\ov k_1} +\cdots\ ,
\quad (D_1)_{ab} = \delta_{ab}+{1\ov 2 \zeta_1}\frac{(f_1)_{ab}}{k_1}
+\cdots\ ,\quad (f_1)_{ab} =  f_{abc} v_1^c\ ,
\\
& J_{n\pm}^a ={1\ov 2\zeta_n} {\del_\pm v_n^{a}\ov k_n} +\cdots\ ,
\quad (D_n)_{ab} = \delta_{ab}+{1\ov 2 \zeta_n}\frac{(f_n)_{ab}}{k_n}
+\cdots\ ,\quad (f_n)_{ab} =  f_{abc} v_n^c\ ,
\label{orrio}
\end{split}
\end{equation}
In this limit the action \eqn{s-modelfin} becomes
\be
\begin{split}
& S = \sum_{i=2}^{n-1} S_{k_i}(g_i)
\\
&
+ {1\ov  2\pi \zeta_1} \int d^2\s\ \del_+ v_1^a (\mathbb{1} +f_1)^{-1}_{ab}\del_- v_1^b
+ {1\ov  2\pi \zeta_n} \int d^2\s\ \del_+ v_n^a (\mathbb{1} +f_n)^{-1}_{ab}\del_- v_n^b
\\
& + {1\ov \pi}\sum_{i=2}^{n-1} k_i\hat \l_{i1} \int d^2\s\
J_{i+} (\mathbb{1} +f_1)^{-1} \del_-v_1
+  {1\ov \pi}\sum_{i=2}^{n-1} k_i\hat \l_{ni} \int d^2\s\
\del_+v_n  (\mathbb{1} +f_n)^{-1} J_{i-}
\\
&+ {1\ov\pi}  \sum_{i=2}^{n-1} k_i \hat \l_{i1}\hat \l_{ni}
\int d^2\s\ \del_+ v_n (\mathbb{1} +f_{n})^{-1}D_i^T
(\mathbb{1} +f_{1})^{-1}\del_- v_1\ .
\label{nobag}
\end{split}
\ee
Note that Euclidean signature imposes a constraint on the parameters
\be
\label{fjh1}
\zeta_1 > 0\ ,\qquad \zeta_n>0 \ ,\qquad \zeta_1 \sum_{i=2}^{n-1}\hat\l_{i1}^2 < 1\ ,
\qquad \zeta_n \sum_{i=2}^{n-1}\hat \l_{ni}^2 < 1\ .
\ee
This $\s$-model represents the interaction of $(n-2)$ WZW models for a group $G$
with two non-Abelian T-duals of the PCM for the same group which in turn interact among themselves.
We note that the $\s$-model \eqn{nobag} being the limit of \eqn{s-modelfin} is itself
integrable. The original action whose non-Abelian T-dual is \eqn{nobag} is that of interacting PCMs.

\no
An interesting truncation of \eqn{nobag} arises if we take the remaining  levels $k_i\to \infty$ as 
\be
\begin{split}
&
g_i =  \mathbb{I} + i {x_i^a t_a \ov \sqrt{k_i}} + \cdots \ ,\qquad \hat \l_{i1}= {\zeta_{i1}\ov \sqrt{k_i}}\ ,\qquad
 \hat \l_{ni}= {\zeta_{ni}\ov \sqrt{k_i}}\ ,
\\
& k_i\to \infty\ ,\qquad i=2,3,\dots, n-1\ .
\end{split}
\ee 
Then \eqn{nobag} becomes 
\be
\begin{split}
& S ={1\ov 2\pi}  \sum_{i=2}^{n-1} \del_+x^a_i \del_-x^a_i
\\
&\quad
+ {1\ov  2\pi \zeta_1} \int d^2\s\ \del_+ v_1^a (\mathbb{1} +f_1)^{-1}_{ab}\del_- v_1^b
+ {1\ov  2\pi \zeta_n} \int d^2\s\ \del_+ v_n^a (\mathbb{1} +f_n)^{-1}_{ab}\del_- v_n^b
\\
& \quad + {1\ov \pi}\sum_{i=2}^{n-1}  \zeta_{i1} \int d^2\s\
\del_+ x_i (\mathbb{1} +f_1)^{-1} \del_-v_1
+  {1\ov \pi}\sum_{i=2}^{n-1} \zeta_{ni} \int d^2\s\
\del_+v_n  (\mathbb{1} +f_n)^{-1} \del_- x_i
\\
& \quad+ {1\ov\pi}  \sum_{i=2}^{n-1}   \zeta_{i1} \zeta_{ni}
\int d^2\s\ \del_+ v_n (\mathbb{1} +f_{n})^{-1}
(\mathbb{1} +f_{1})^{-1}\del_- v_1\ .
\label{nobag9}
\end{split}
\ee
This represents the mutual interactions of $(n\!-\!2)\dim G$ free fields with the two non-Abelian T-duals of PCM for a group 
$G$. We may clearly simply even further by consistently taking $\zeta_{i1}=0$ or $\zeta_{ni}=0$.

\section{Discussion and future directions}

We have construct the all loop effective action
of a general class of models whose UV Lagrangian is the sum of an arbitrary number $n$ of WZW models based on the same group $G$, but at different levels. Although the complete effective action can be quite involved, at the linear level the theory is driven away from the conformal point by operators bilinear in the WZW currents. These current bilinears involve currents belonging to both the same and different CFTs. 
 Hence we have self- as well as mutual interactions of current algebra theories in their most general form.
In general our models depend on $n^2$ general coupling matrices. We considered a consistent truncation of our models that depends on $2(n-1)$ couplings and showed that  the theory is classically integrable by finding the appropriate Lax connection. Turning on even one more coupling will generate other couplings through quantum corrections and the theory will most likely cease to be integrable. Subsequently, we proved that the theory possesses  certain  non-perturbative symmetries in the space of couplings $\hat \l_{i1}$ and $\hat \l{ni}$ and calculated the exact in the deformation parameters $\b$-functions of our models which can be cast in a particularly simple and compact form.
This fact allowed us to fully determine and classify the fixed points of the RG flow in particular those that
are IR stable. Last but not least we consider non-Abelian T-duality type limits in the case when the theory is integrable.

A number of open questions remain to be addressed. Given that our models provide concrete realizations of integrable flows between exact CFTs it would be interesting to elucidate the nature and symmetries of the corresponding IR stable CFTs. Furthermore, one could exploit the aforementioned non-perturbative symmetries that our models enjoy to compute the anomalous dimensions of current operators, as well as that of primary operators in a similar manner to that in\cite{Georgiou:2015nka,Georgiou:2016iom,Georgiou:2016zyo,Georgiou:2017aei,c-function:2018,Georgiou:2017oly}.  One could also calculate the exact in the deformation parameters $C$-function of the models presented here as was done in \cite{c-function:2018} for simpler cases.  In that respect the general results in \cite{c-function:2018} should be a useful
starting point.

Another direction would be to consider the case where more of the couplings are non-zero, as well as the case where each of the WZW models is based on a different group. It is notable that  in the latter case all formulae of section 2 will still be valid
 after slight modifications. In addition, one could search for integrable deformations in these more general cases. Compared to the integrable models presented in this work, we expect an even richer structure of the RG equations to be unveiled. Recently, a class of integrable models consisting of $N$ coupled principal chiral models each with a WZW term and based on the same group $G$ was presented in  \cite{Delduc:2018hty}. The construction was based on
an association of integrable field theories with affine Gaudin models having an arbitrary number of  sites.
It would be interesting to see if these models bare any relation to the ones constructed in this work (see also \cite{Georgiou:2018hpd}) or to the models with $N$ sites firstly presented in \cite{Georgiou:2017oly}.
Finally, one could try to embed our models to solutions of type-IIB or type-IIA supergravity {\color{red} and/or} construct the corresponding $\eta$-deformed integrable models.

\section*{Acknowledgments}

The work of G.G. on this project has received funding from the Hellenic Foundation for Research and Innovation
(HFRI) and the General Secretariat for Research and Technology (GSRT), under grant
agreement No 234.
K. S. would like to thank the Theoretical Physics Department of CERN for hospitality and
financial support during part of this research.

\end{document}